# A High Frequency Active Clamp Forward Converter with Coreless Transformer

Reza Asrar Ghaderloo
ECE Department
*University of Maryland at college Park*
Maryland, United States
rezasrar@umd.edu

Ali Parsa Sirat
ECE Department
*University of North Carolina at Charlotte*
Charlotte, United States
aparsasi@charlotte.edu

Abbas Shoulaie
ECE Department
*Iran University of Science and Technology*
Tehran, Iran
shoulaie@iust.ac.ir

*Abstract*—In this paper, a highly compact, low power (< 10W), high frequency (2 MHz) isolated active clamp forward converter, comprising a coreless Printed Circuit Board-based transformer is proposed. To decrease the size of converter, high switching frequency is considered which lead to decrease in inductor, capacitor and transformer size. Highly switch loss due to hard switching is an important constraint of forward topology to increase frequency. In this paper, the active clamp circuit is added to forward topology to achieve zero voltage switching and decrease switching loss drastically. Due to zero voltage switching, the proposed converter can operate in high frequency. The principle of active clamp forward converter is described in this paper. Another constraint to increase the switching frequency of forward converter is transformer core losses. In this paper, coreless PCB-based transformer is proposed and implemented to be utilized in the structure of the active clamp forward converter. Instead of classic core-based transformer, using a PCB-based transformer as the power transmitter has increased the efficiency due to elimination of core hysteresis loss. The equivalent circuit, transfer function and input impedance of PCB-based coreless transformer are presented in high frequency. Finally, an experimental prototype of the active clamp forward converter which uses a coreless transformer is implemented. The experimental results of proposed converter are presented to evaluate the theoretical analysis and performance.

I. INTRODUCTION

In the 21st century, the demand for small electronic equipment has increased for use in daily life. Therefore, one of the primary goals for power electronics engineers is to minimize the size and the cost of switching mode power supplies (SMPS). The increase of the switching frequency is the most common approach to minimize the size of a converter as it leads to size decrease of passive components like capacitors, inductors, and transformers. However, the increasing of the switching frequency raises various concerns such as hysteresis and switching loss.

In this paper, an isolated, high frequency, low power (< $10W$) DC-DC converter based on the forward topology is proposed.

One of the most common low-cost low power DC-DC topologies is the forward converter. Fewer component requirements and the size of forward converters make them an excellent choice of topology for a DC-DC isolated converter. Increasing switching frequency to minimize the size of a forward converter leads to two significant challenges (i) the increase in switching losses due to hard switching transition and (ii) the increase in transformer hysteresis and eddy current losses.

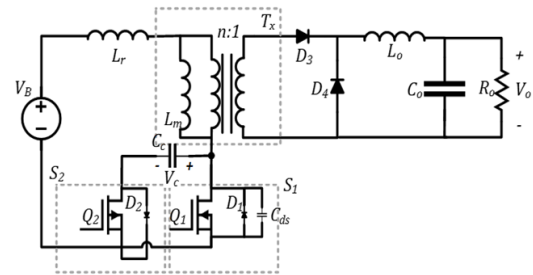

Fig. 1. Low side active clamp forward converter

The approach to tackle the hard switching losses in forward converter is to implement an auxiliary active clamp circuit. As show in 1, the active clamp circuit included two components, which results in zero voltage switching (ZVS) behavior in the forward converter. The main drawback of the active clamp forward converter (ACFC) compared to the conventional forward topology is an additional MOSFET and capacitor that can increase the cost and size of the converter. The ACFC is categorized in two groups of high-side and low-side. High side and low side ACFC are studied in [1] and [2], respectively.

Researchers tend to use both low-side and high-side ACFC, however, to avoid the complexity of high side switch drive, the low-side ACFC is chosen to investigate in this paper. The voltage drop in output diode of ACFC increase losses and decrease the efficiency of converter. To overcome this drawback, MOSFET synchronous rectification is used to minimize the voltage drop [3,4] . In mentioned papers, the core-based transformer is used which can result in lower efficiency in high frequency. The hysteresis and eddy current losses of ferrite core-based transformers increase significantly in high frequencies. Implementation of a coreless PCB-based transformer helps overcome the eddy and hysteresis losses associated with the ferrite core. Moreover, coreless transformers result in minimizing the converter size.

The winding configuration of the proposed coreless transformer is depicted in Fig. 2. The primary and secondary windings are formed by using top and bottom layers of a PCB in a way that their center coincide together. The equivalent circuit of coreless transformer is described in [5]. The resulting capacitance between windings, mutual inductance and leakage inductance are modeled. In coreless transformer design, the inductance depends on both the number of turns and the geometry of the windings. The effect of different geometrical characteristics on coreless transformer inductance is studied in

[6]. Indeed, the effect of copper thickness, clearance between traces, PCB thickness and variety of geometry effects on transformer characteristics are studied. The input impedance and transfer function of coreless transformers is also studied in [7]. It has resulted in investigation of coreless PCB-based transformer behavior in high frequencies.

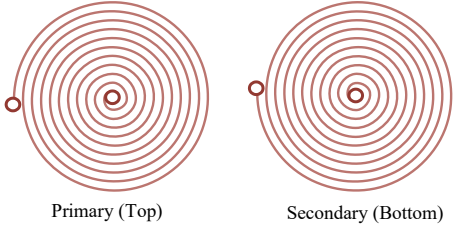

Fig. 2. Coreless transformer

A coreless transformer is utilized in [8] and [9] to perform as the energy transmitter in an LLC half-bridge resonant converter. Furthermore, to improve the coreless transformer power transfer capacity, a compensator capacitor is added to the primary and secondary windings in wireless battery charger applications [10].

In this paper, the implementation of high frequency forward topology is investigated. Due to tackle the hard switching transition issues, the active clamp circuit is added to topology. Moreover, to mitigate the core-based transformer losses in high frequencies, a coreless PCB-based transformer is utilized as power transformer.

This paper is organized as follows. In Section II, a brief description of ACFC operation is provided and different conditions which are required for soft switching transition are identified. In Section III, the equivalent circuit of a coreless PCB-based transformer is illustrated to calculate the transfer function and the input impedance. Then, the implemented coreless transformer characteristics are presented. In Section IV, experimental results of the laboratory prototype of ACFC, which utilized coreless transformer, are used to verify theoretical predictions. Section V concludes the paper.

## II. ACTIVE CLAMP FORWARD CONVERTER

The circuit topology of the ACFC converter is depicted in Fig. 1. In this figure, $S_1$ and $S_2$ represent the switch blocks of ACFC. Each switch block is comprised of a MOSFET ($Q_1$, $Q_2$) and its anti-parallel diode ($D_1$, $D_2$). The isolation transformer is represented by $T_x$, and $D_3$ and $D_4$ are the forward diode and freewheeling diode, respectively. The output capacitor and output inductor are represented by $C_c$ and $L_o$, respectively. The additional switch $S_2$ and the clamping capacitor $C_c$ are used to replace the anti-saturation winding in conventional forward converters, which can reset the residual transformer flux. In this structure, the input voltage source stores the residual energy of the inductor $L_m$. Thanks to the additional active clamp circuit, both switches can be soft switched under zero voltage crossing, which results in decreased switching losses and increased efficiency. To make the analysis simpler, the following assumptions are made:

1) The intrinsic capacitor $C_{ds}$ is negligible compared to the capacitance of $C_c$

2) To consider constant values for the output current and voltage, the output capacitor and inductor are assumed to be large.

3) The short dead-time between two switches is neglected. Therefore, the conduction time for $S_1$ and $S_2$ can be expressed as $DT_S$ and $(1-D)T_S$, respectively. The switching period is represented by $T_S$.

In case which switch $S_1$ is on and $S_2$ is off, the voltage across the primary winding of transformer is equal to input voltage. Moreover, in case which switch $S_1$ is off and $S_2$ is on, the voltage across primary winding is equal to $V_C - V_B$, where the input voltage and the clamping capacitor voltage are represented by $V_B$ and $V_C$, respectively. Considering the average voltage computation across the primary winding of transformer, the expression below can be written:

$$V_B DT_s = (V_C - V_B)(1-D)T_s \quad (1)$$

The active clamping capacitor voltage can be written in terms of the input voltage as:

$$V_C = \frac{1}{1-D} V_B \quad (2)$$

Equation 2 illustrates the performance of a boost converter with the input voltage of $V_B$ and the output voltage of $V_C$. When the switch $S_1$ is off and $S_2$ in on, the $S_1$ voltage is equal to the clamping capacitor voltage and can be written as:

$$V_{ds1} = V_C = \frac{1}{1-D} V_B \quad (3)$$

Fig. 3 shows nine different operating modes and waveforms of an ACFC during one switching period. The equivalent circuit corresponding for each operating point is depicted in Fig. 4.

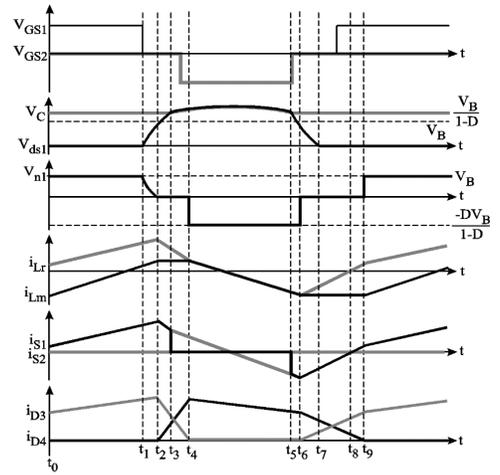

Fig. 3. Theoretical waveform of the ACFC

Mode 1 ($t_0 \sim t_1$): At t=$t_0$, the $S_1$ gate signal is applied to turn the switch on, and the $S_2$ gate signal is removed to turn off the switch. The magnetizing current ($I_{Lm}$) linearly increases since primary winding voltage, $V_{n1}$, is fixed and equal to the input voltage ($V_B$). The transformer secondary winding voltage, $V_{n2}$, is $V_B/n$, thus $D_4$ is turned off and $D_3$ is turned on. The output

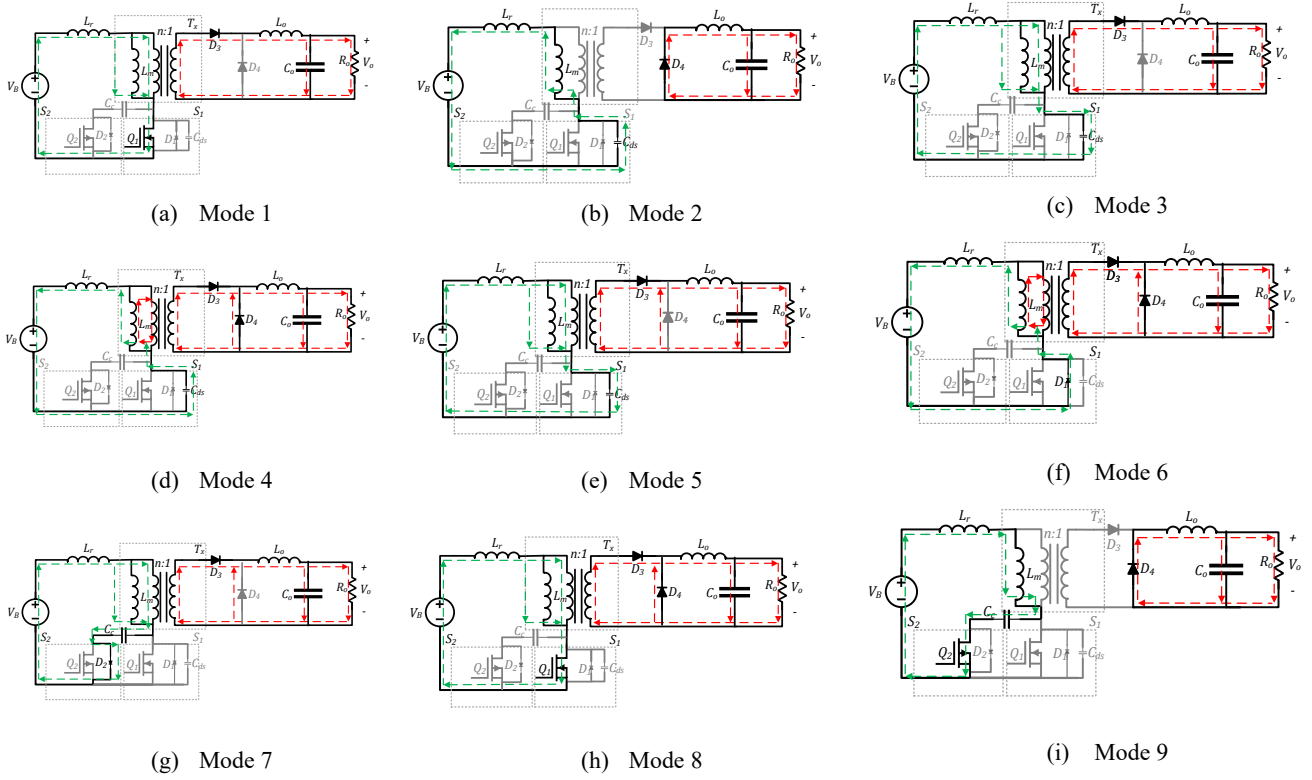

Fig. 4. Equivalent circuit of operation modes

inductor $L_0$ gets energy by the input voltage via the diode $D_3$ and the transformer.

Mode 2 ($t_1 \sim t_2$): at the start of this mode $S_1$ is turned off. The $S_1$ switch intrinsic capacitor, $C_{ds1}$, gets charged by the inductor reflected current, which leads to the increase of $V_{ds1}$ and decrease of $V_{n1}$. When $V_{ds1}$ gets equal to $V_B$, this mode ends.

Mode 3 ($t_2 \sim t_3$): At t=$t_2$, $V_{n1}$ is equal to zero, hence $D_3$ and $D_4$ will be turned on. The leakage inductance current decreases because the diode current $I_{D3}$ reduces. In this mode, $I_{Lm}$ stays unchanged. When $V_{ds1}$ increases to $V_B/(1-D)$, this mode ends.

Mode 4 ($t_3 \sim t_4$): At the beginning of this mode, $S_1$ is turned off. To transfer energy to the clamping capacitor $C_c$, the primary side current is going through $D_2$. Hence, the ZVS transition circumstance is provided for switch $S_2$ and the gate signal corresponding to $S_2$ must be applied. When the current of $D_3$ goes to zero, this mode ends.

Mode 5 ($t_4 \sim t_5$): At t=$t_4$, the current $I_{D3}$ is zero, consequently, $L_M$ and $L_r$ are in series structure and engage in the resonance with $C_c$. The voltage $V_{n1}$ is equal to $DV_B/(1-D)$ with reverse polarity. In this condition, $I_{Lr}$ and $I_{Lm}$ are equal and remain descending. As the direction of $I_{Lr}$ reverses, the existing energy in $L_m$ is transferred to the input source. By removing the gate signal of $S_2$, this mode ends.

Mode 6 ($t_5 \sim t_6$): At t=$t_5$, the switch $S_2$ will be in the off position, therefore, the intrinsic capacitor $C_{ds}$ gets discharged since the current $I_{Lr}$ flows through that. When the voltage $V_{ds1}$ descends to $V_B$, this mode ends.

Mode 7 ($t_6 \sim t_7$): At t=$t_6$, the voltage $V_{ds1}$ and $V_B$ are equal, and the voltage $V_{n1}$ changes to zero once again. Both current $I_{D3}$ and $I_{Lr}$ increase in this mode. The value of the current $I_{Lm}$ remains constant. When the voltage $V_{ds1}$ reaches zero, this mode ends.

Mode 8 ($t_7 \sim t_8$): At t=$t_7$, the inductor current $I_{Lr}$ sinks back to the input voltage source through the anti-parallel diode $D_1$. Therefore, the zero-voltage transition is provided for $S_1$. During this mode, the gate signal of $S_1$ must be applied. With the slope of $V_B/L_r$, the current $I_{Lr}$ starts increasing. When $I_{Lr}$ gets to zero, this mode ends.

Mode 9 ($t_8 \sim t_9$): At t=$t_8$, the current $I_{Lr}$ goes back to positive value and the current $I_{D4}$ keeps reducing. As $I_{D4}$ reaches zero and $D_4$ turns off, this mode ends and next switching cycle starts.

## III. CORELESS TRANSFORMER

The magnetic core loss is a significant constraint in high switching frequencies. Therefore, the proposed coreless transformer is an appropriate alternative. In addition, it can reduce the cost and size of the converter significantly. The coreless transformer geometric structure has a considerable influence on the values of the equivalent circuit component. Fig. 5 shows the geometric structure of a coreless transformer. The PCB board thickness, winding radius, track height, track separation and track width are defined in the Fig. 5.

### A. Equivalent Circuit

To study the equivalent circuit of a coreless transformer in high frequencies, the capacitance between windings should be considered. In addition, the resistor which presents core losses in classic models was omitted in the equivalent circuit of coreless transformers. The equivalent circuit of a coreless transformer is depicted in Fig. 6. Where $R_1$ is the primary winding resistance, $L_{lk1}$ is the primary winding leakage inductance, $C_1$ is the primary winding capacitance, $R'_2$ is the secondary winding resistance referred to primary, $L'_{lk2}$ is the secondary winding leakage inductance referred to primary, $C'_2$ is the secondary winding capacitance referred to primary, $C'_{12}$ is the capacitance between primary and secondary winding, and $L_{m1}$ is the mutual inductance.

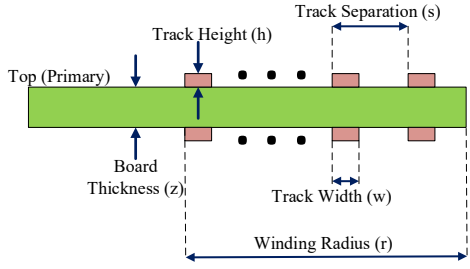

Fig. 5. Geometric structure of coreless transformer

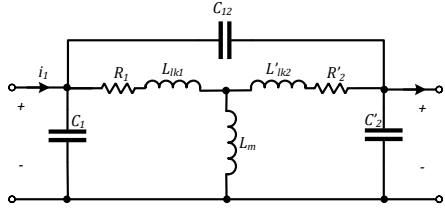

Fig. 6. High frequency model for PCB transformer

The transfer function and the input impedance of a coreless transformer can be written, based on the circuit model of Fig. 6, as:

$$\frac{V_s}{V_p} = \frac{1}{n} \frac{\frac{1}{R_1+sL_{lk1}} + sC'_{12}\left[(R'_2+sL'_{lk2})\left(\frac{1}{R_1+sL_{lk1}}+\frac{1}{sL_{M1}}\right)+1\right]}{\frac{1}{R'_2+sL'_{lk2}}+\left[\left(\frac{1}{R'_2+sL'_{lk2}}+sC'_{12}+sC_2\right)\right]} \times \frac{1}{\left[(R'_2+sL'_{lk2})\left(\frac{1}{R_1+sL_{lk1}}+\frac{1}{sL_{M1}}\right)+1\right]} \quad (4)$$

$$Z_{in} = \frac{1}{sC'_1 + sC'_{12}\left(1-\frac{V_s}{V_p}\right) + (1-W)/(R_1+sL_{lk1})} \quad (5)$$

Where:

$$W = \frac{sC'_{12} + \frac{R'_2+sL'_{lk2}}{R_1+sL_{lk1}}\left(\frac{1}{R'_2+sL'_{lk2}}+sC'_{12}+sC_2\right)}{\left[(R'_2+sL'_{lk2})\left(\frac{1}{R'_2+sL'_{lk2}}+\frac{1}{sL_m}\right)+1\right]} \times \frac{1}{\left(\frac{1}{R'_2+sL'_{lk2}}+sC'_{12}+sC_2\right) - \frac{1}{R'_2+sL'_{lk2}}} \quad (6)$$

To be used in the structure of an ACFC, a coreless transformer with geometric dimensions, projected in Table I, is designed and fabricated. After implementation of the coreless PCB-based transformer, the equivalent circuit component values are measured and shown in the Table I.

### B. Input Impedance and Transfer Function

With equivalent circuit component values is hand, the equations (4) and (5) are implemented in the MATLAB environment. The transfer function gain and the input impedance curve, with respect to the frequency, are depicted in Fig. 7 and Fig. 8, respectively.

As illustrated in Fig. 7, the transfer function magnitude was approximately equal to 0.72 for frequencies smaller than 6.5 MHz, and then it decreases. As shown in Fig. 7, the bandwidth of transfer function was from a few kilohertz to 10 MHz.

Table I. Equivalent circuit component values of designed coreless transformer

| Geometric parameters | | Component values | |
|---|---|---|---|
| Inner radius ($r_1$) | 2 mm | $C_1$ | 4pF |
| Outer radius ($r_2$) | 25 mm | $C_2$ | 4pF |
| Conductor width (a) | 0.635 mm | $C_{12}$ | 16pF |
| Conductor separation (s) | 0.47 mm | $L_{lk1} = L_{lk2}$ | 3.9 µH |
| Pri./Sec. winding (N) | 23/23 | $L_m$ | 10.1µH |
| Conductor height (h) | 35 u | $R_1$ | 1.27Ω |
| Board thickness (z) | 1.6 mm | $R_2$ | 1.27Ω |

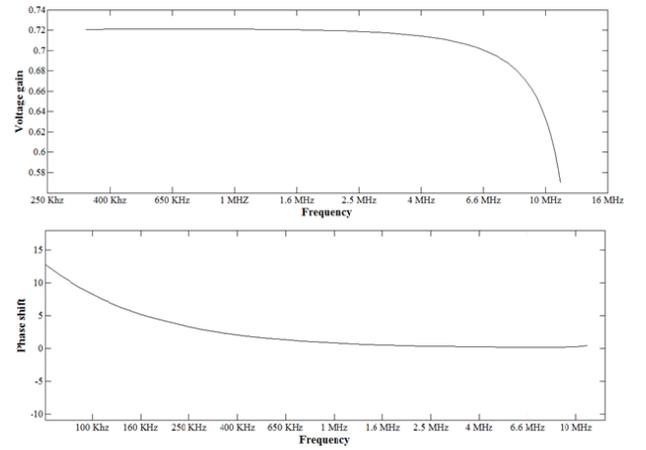

Fig. 7. Transfer function of designed coreless transformer

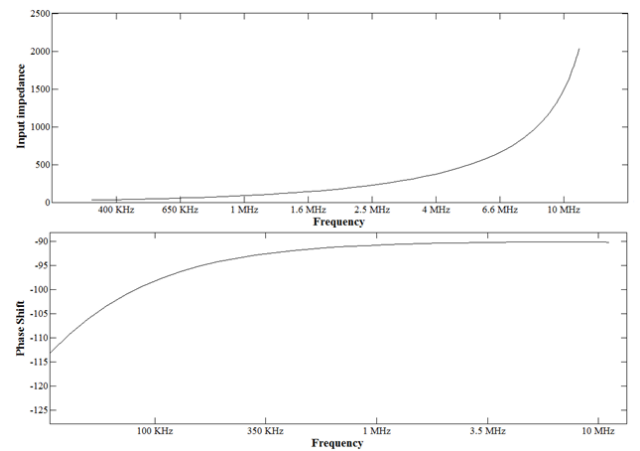

Fig. 8. Input impedance of designed coreless transformer

Fig. 8 shows the input impedance increased with respect to the frequency. Thereby, the increase in frequency leads to the reduction of the input current. Consequently, the coreless transformer is more efficient higher frequencies which results in improvement of the converters power efficiency. However, it is obvious that the increase in frequency is subjected to other constraints like the MOSFET switching speed due to its intrinsic capacitance, the output diode switching losses, the AC resistance of a coreless transformer, the output inductor, and the PCB tracks.

## IV. SWITCHING FREQUENCY SELECTION PROCEDURE

Based on the ACFC operation principle and the coreless transformer characteristics, the proposed ACFC with coreless transformer is designed and implemented.

To guarantee that switch $S_1$ operates in ZVS, the minimum value for the inductance of $L_r$ at $t = t_6$ should be selected in a way that its energy becomes more than energy stored in $C_{ds}$ capacitor. Thereby, it can be written as:

$$L_r > \frac{C_{ds}V_{in}^2}{|i_{Lr}(t_6)|^2} \qquad (7)$$

According to ACFC operating principle, the current $i_{Lr}$ at $t = t_6$ can be computed as:

$$i_{Lr}(t_6) = \frac{DV_{in}}{2L_m f_s} \qquad (8)$$

From (7) and (8):

$$L_r > \frac{4L_m^2 f_s^2 C_{ds}}{D^2} \rightarrow f_s < \sqrt{\frac{L_r D^2}{4L_m^2 C_{ds}}} \qquad (9)$$

The intrinsic capacitor $C_{ds}$ of main switch IRF510 is 80 pF, duty cycle is set to 0.5 and inductance $L_r$ and $L_m$ of designed coreless transformer is 3.9µH and 10.1µH, respectively. Therefore, the switching frequency should be below 5.6 MHz to achieve ZVS operation in the proposed converter based on (9). Therefore, the switching frequency is increased from 1.1 MHz to 1.4 MHz, then 1.9 MHz and finally 2.5 MHz to provide ZVS transition for switches and show the coreless transformer efficiency improvement in higher frequencies.

Conversely, to optimize the voltage stress on the main and auxiliary switches, the resonance period of $C_c$ and inductance $(L_r + L_m)$ in mode 5 and 6 of detailed converter must be at least 10 times greater than the turn-off time of main switch $S_1$. It can be expressed as:

$$C_c \geq \frac{25(1-D)^2}{\pi^2 f_s^2 (L_m + L_r)} \qquad (10)$$

From (10) it can be written:

$$f_s \geq \sqrt{\frac{25(1-D)^2}{\pi^2 C_c (L_m + L_r)}} \qquad (11)$$

As the clamp capacitor is set to 1µF, the switching frequency should be greater than 650 KHz. Therefore, the chosen switching frequencies (1.1 MHz, 1.4 MHz, 1.9 MHz and 2.5 MHz) meet all constraints.

## V. EXPERIMENTAL RESULTS

To experimentally verify the operation of active clamp forward converter using coreless transformer, the converter shown in Fig. 9 is implemented. In this converter, the input voltage and duty cycle are set to 40V and 0.5, respectively. The selected key component of circuit and their values are tabulated in table III.

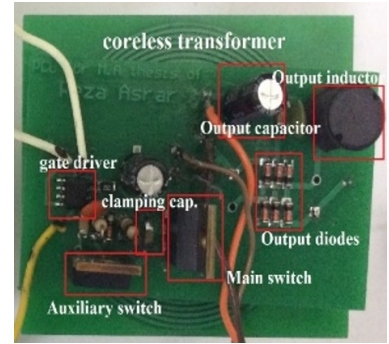

Fig. 9. Illustrative prototype of proposed ACFC with coreless transformer

Table II. Calculated components of prototype

| Main switch, $S_1$ | IRF510 (100V/5.6A) |
|---|---|
| Auxiliary switch, $S_2$ | IRF9630 (200V/6.5A) |
| Diode $D_3, D_4$ | 3 paralleled 1N4148(100V/450mA) |
| Leakage inductance, $L_r$ | 3.9 µH |
| Magnetizing inductance, $L_m$ | 10.1 µH |
| Clamping capacitor, $C_c$ | 1 µF |
| Output inductor, $L_o$ | 100 µH |
| Output capacitor, $C_o$ | 27 µF |

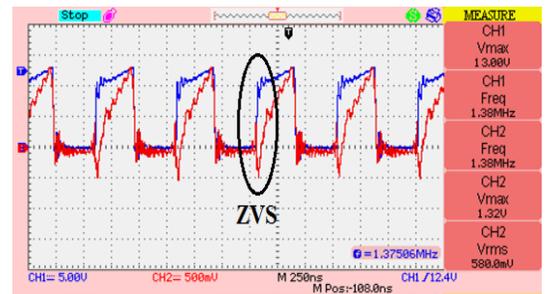

Fig.10. Measured waveform of main switch current (red) and driving signal $V_{GS1}$ (blue)

The main switch $S_1$ current and its gate signal are shown in Fig.10 to verify ZVS soft switching transition of $S_1$. As expected, it is evident that negative current flowed through the body diode of switch before applying its driving signal $V_{GS1}$. Thus, this switch is operating in ZVS switching transitions.

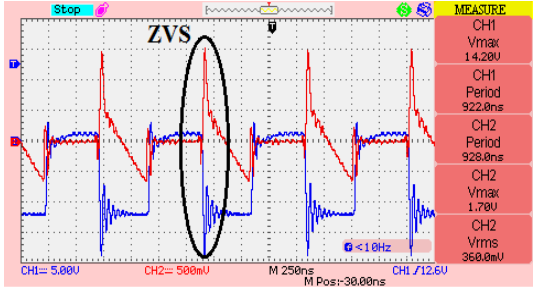

Fig. 11. Measured waveform of auxiliary switch current (red) and driving signal $V_{GS2}$ (blue)

Fig. 11 shows the measured waveform of auxiliary switch $S_1$ current and its driving signal $V_{GS2}$. The auxiliary switch is a p-channel MOSFET, thus, negative gate voltage turns it on. Similar to the main switch, ZVS soft switching transition of $S_2$ is obvious in Fig. 11.

Table IV. Efficiency of designed coreless transformer

| Frequency (MHz) | 1.1 | 1.4 | 1.9 | 2.5 |
|---|---|---|---|---|
| Efficiency | 72 % | 79 % | 88 % | 92 % |

Table III. Efficiency of designed coreless transformer

| Frequency (MHz) | 1.1 | 1.4 | 1.9 | 2.5 |
|---|---|---|---|---|
| Efficiency | 72 % | 79 % | 88 % | 92 % |

The efficiency of the implemented converter and coreless transformer in different switching frequency are listed in Table-III and Table IV, respectively. To calculate the coreless transformer efficiency, the pure power delivered to primary and secondary windings are calculated. To calculate the delivered power to the primary winding, the gate driving and on-state losses of switches are subtracted from the input power. Similarly, the delivered power by the secondary winding is approximately equal to the sum of the output power and the loss of diodes and inductor. According to Table III and Table IV, the increase in switching frequency can improve the coreless transformer efficiency, However, due to the increase in loss of output diodes and AC resistor of coreless transformer and the PCB tracks, the increase in frequency do not necessarily improve the converter efficiency in frequencies above 1.9 MHz.

## VI. CONCLUSION

In this paper, a high-frequency, low-power active clamp forward converter with a coreless transformer is proposed. The active clamp additional circuit can provide ZVS switching condition for the forward converter. Furthermore, using a coreless transformer in an active clamp forward converter can lead to decrease in hysteresis and eddy current losses in frequencies within a few megahertz. Additionally, a coreless transformer can decrease the size of the converter. The operational diagrams of a coreless transformer and ACFC are described. Eventually, to evaluate the effectiveness of theoretical computations, the experimental results based on a laboratory prototype are provided.